\documentclass[number,3p,a4paper,twocolumn, preprint]{elsarticle}
\usepackage[strings]{underscore}
\usepackage[pdfauthor=author,colorlinks=false,hidelinks]{hyperref}
\usepackage{amsmath}
\usepackage{amssymb}
\usepackage{cleveref}
\usepackage{xcolor}
\usepackage[scientific-notation=true]{siunitx}
\usepackage{subcaption}
\usepackage{placeins}
\usepackage{float}
\pdfoutput=1

\newcommand{\ra}{\operatorname{Ra}}
\newcommand{\pr}{\operatorname{Pr}}
\newcommand{\ste}{\operatorname{St}}

\newcommand{\phist}{\phi^\star}

\journal{Thermal Science and Engineering Progress}
\begin{document}
\begin{frontmatter}

\title{Heat transfer modulation in Phase Change Materials via fin insertion}
\author[1]{Paolo Proia\corref{cor1}}
\ead{paolo.proia@uniroma2.it}
\affiliation[1]{Department of Enterprise Engineering ``Mario Lucertini'', Tor Vergata University of Rome, Via del Politecnico 1, 00133 Rome, Italy}
\cortext[cor1]{corresponding author}

\author[2]{Mauro Sbragaglia}
 \ead{sbragaglia@roma2.infn.it}
\affiliation[2]{Department of Physics & INFN, Tor Vergata University of Rome, Via della Ricerca Scientifica 1, 00133, Rome, Italy.}

\author[1,3]{Giacomo Falcucci}
\ead{giacomo.falcucci@uniroma2.it}
\affiliation[3]{Department of Physics, Harvard University,  33 Oxford Street, 02138 Cambridge, Massachusetts, USA.}

\date{\today}
\begin{abstract}
We leverage a large set of numerical simulations to study optimized geometrical configurations for Phase Change Materials (PCMs) cells. We consider a PCM cell as a square enclosure with a solid substance that undergoes melting under the effect of a heat source from one side and under the effects of buoyancy forces. Moreover, an additional source fin with prescribed length $l$ and height $h$ protrudes into the cell perpendicularly from the heat source. The fin prompts enhanced heat transfer and convection within the PCM cell, thus shortening (in comparison to a finless cell) the melting time $t_m$ needed for all the PCM material to melt and transit from the solid to the liquid phase. This improvement is systematically studied as a function of the fin geometrical details  ($l$, $h$), as well as the Rayleigh number $\ra$ -- encoding the importance of buoyancy forces with respect to diffusion/dissipation effects -- and the Stefan number $\ste$ -- encoding the importance of sensible heat with respect to latent heat. Overall, our systematic study in terms of the free parameters $l$, $h$, $\ra$ and $\ste$ offers inspiring insights to optimize the structure of a PCM cell during its manufacturing process and suggests optimal operating conditions for such geometrical configurations.
\end{abstract}


\end{frontmatter}
\section{Introduction}
The exploitation of storage and buffer technologies is required to mitigate the inconsistency of renewable energy sources~\cite{kittner-energy-2017}. Among the different storage solutions, thermal energy storage via phase change materials (PCMs) is gaining increasing attention in recent years~\cite{sadeghi-energy-2022}. PCMs are characterized by large latent heat, allowing the storage of considerable amounts of thermal energy at a given temperature value, which can be fixed \textit{a priori} by choosing the appropriate material and, thus, its thermal properties. Thermal energy storage (TES) systems are widely employed in many sectors, from buildings to solar plants, from battery management to hydrogen-based energy systems, since heat recycling is a key topic in many aspects of the ecological transition~\cite{cabeza-materials-2011,chavan-comprehensive-2022,altuntas-comprehensive-2025,togun-critical-2024,wang-critical-2022,faraj-review-2021,lin-review-2018,nivedhitha-advances-2024,lin-review-2021}.\\
The main drawback in the exploitation of PCMs for TES lies in their limited thermal conductivity, leading to poor power performance, which has prompted a vast scientific investigation for the different strategies to overcome such a limitation during the manufacturing process~\cite{wu-Thermal-2020,togun-critical-2024,ma-review-2023,wu-phase-2021,michalrogowski-recent-2023,zhao-review-2022,lin-review-2018}, including (but not limited to) modifying the enclosure shape and/or boundaries~\cite{rashid-review-2022,ma-review-2023}, the insertion of fins~\cite{zhu-heat-2022,mostafavi-Thermal-2022,jourabian-simulation-2012,bhowmik-embedding-2025}, the addition of nanoparticles~\cite{bouzennada-numerical-2023,sachan-characterization-2025}, the incorporation of metal foams~\cite{gao-study-2010,pourfallah-experimental-2025} or combining two or more of these techniques~\cite{gao-lattice-2017,togun-critical-2024,jaberi-optimizing-2025}. Given both the complex theoretical and practical aspects of heat transfer, PCM optimization proves to be an interesting topic of research. Among the many optimization routes, in this paper we specifically focus on the insertion of a single source-like fin inside an enclosure filled by a solid substance that undergoes melting. The use of fins to enhance thermal properties is a well-explored field~\cite{ma-review-2023,wu-phase-2021,lin-review-2018,choudhari-Numerical-2020,gao-lattice-2017,jourabian-simulation-2012,oskouei-closecontact-2024,sharifi-enhancement-2011,zhu-heat-2022,mostafavi-Thermal-2022,xie-investigation-2025,bhowmik-embedding-2025,jaberi-optimizing-2025,laasri-investigation-2022,arshad-transient-2020}, as they provide a simple yet effective way of improving heat transfer and melting rates. For example, in~\cite{wu-phase-2021} PCM cells with a conductive fin were studied, showing that the melting time is reduced by increasing the fin length and lowering the fin height simultaneously. Gao et al.~\cite{gao-lattice-2017} investigated the combined effects of a porous medium and a fin, finding that the melting speed increases with the length of the fin and a decrease of the heat capacity of the fin. In~\cite{jourabian-simulation-2012}, the study on the position and size of a single fin was conducted, comparing the melting rate and finding only the length of the fin to impact on melting speed. The study conducted in~\cite{sharifi-enhancement-2011} focused on the simulation of a multi-finned PCM system and the subsequent development of an algorithm to estimate melting rates. The authors found that the insertion of multiple fins speeds up the initial part of the melting phase, but once the inter-fin substance is molten, the melting slows down. The authors relate the transition between the two regimes to the fin length. The role of convection with respect to other heat transfer mechanisms is investigated in~\cite{oskouei-closecontact-2024}: the authors focus on close contact melting, related to the density difference between molten and solid PCM. As the solid phase falls down for gravity, the molten part moves to its sides, keeping a near-constant film thickness between the heated surface and the solid PCM, boosting the thermal transfer. In a configuration with multiple close fins, close contact melting provides a considerable increase in the overall liquefaction. Expanding on the topic of fin shape optimization the authors in~\cite{zhu-heat-2022} develop a topology optimization method and study the effect of fin number, position, shape and branching structure, finding non-uniform arrays of multi branched fins to be the most beneficial arrangement.\\
In general, the presence of fins is expected to influence the behavior of the system, as their position and size promote the development of convective fluid motions, which play a fundamental role in heat transfer and melting~\cite{yang-morphology-2023,jany-scaling-1988,du-sea-2023,wang-Numerical-2017}. Convection is indeed a quite complex phenomenon and the occurrence of melting/solidification triggers highly non-linear behaviors~\cite{wang-equilibrium-2021,du-physics-2024,huber-lattice-2008} which are dependent on the fluid turbulence and motion~\cite{wang-ice-2021,wang-how-2021,rabbanipouresfahani-basal-2018}, on the substance properties (e.g., its salinity~\cite{du-sea-2023}), on the relative angle between the temperature gradient and gravity~\cite{wang-ice-2021,korti-experimental-2020,kasper-numerical-2021}, and, of course, on the boundary conditions~\cite{tisserand-comparison-2011,qin-numerical-2021,freund-rayleigh-2011,rashid-review-2022,toppaladoddi-roughness-2017,wang-thermal-2017}.

Our aim is to conduct a parametric investigation on the position and size of a single fin, quantifying the cell total melting time and its relation to the formation and structure of convective plumes.
The system dynamics will be studied as a function of the characteristic non-dimensional Rayleigh ($\ra$) and Stefan ($\ste$) numbers~\cite{jany-scaling-1988,huber-lattice-2008}.  The results generally confirm that convection-favoring configurations are the most efficient in terms of heat transfer and melting velocity. Interestingly, we show that the optimal geometrical configuration in terms of fin length and height, characterized by the minimal melting time, non-trivially depends on the choice of $\ra$ and $\ste$. \\
The paper is organized as follows: in~\cref{sec:problem} we present the mathematical statement of the problem and discuss the system set-up; in~\cref{sec:results} results and discussion are presented; conclusions will follow in~\cref{sec:conclusions}.

\section{Problem Statement and set-up}\label{sec:problem}
The structure of the PCM cell studied in this work is shown in~\cref{fig:cell-geometry}: the cell is a 2D square with a side length of $L$, with a heat source at temperature $T_H$ on the left side and insulated on the other three sides ($\partial_{\vec{n}}T=0$ where $\vec{n}$ is normal to the wall). It is filled with a PCM at its melting temperature $T_C$. A fin protrudes perpendicularly from the heat source. It has a fixed surface $S$ and is treated as a source (that is, there is no conduction inside the fin). It is positioned at height $h$ and its length is $l$; both these dimensions are normalized to $L$ as $\hat{l}=l/L$ and $\hat{h}=h/L$. We conducted different numerical simulations by varying $(\hat{h},\hat{l})$ to study the influence of the fin structure on the characteristic melting time. The mathematical model for the melting process is based on the continuity and Navier-Stokes equations supplemented with an advection diffusion equation for the temperature field featuring a melting term~\cite{huber-lattice-2008,proia-melting-2024}:
\begin{gather}
    \partial_t\rho+\vec\nabla \cdot \left(\rho \vec{u}\right)=0\label{eq:ns-mass}\\
            (\partial_t+\vec{u}\cdot\vec\nabla)\vec{u}=\nu\nabla^2\vec{u}-\frac{\vec\nabla p}{\rho}- \alpha(T-T_0)\vec{g}\label{eq:ns-momentum}\\
\partial_t T +\vec{\nabla}\cdot(T \vec{u})= \kappa\nabla^2T - \frac{L_f}{C_p}\partial_t\phi   \label{eq:temperature-equation}
\end{gather}
where $\rho=\rho(\vec{x},t)$ is the density of the fluid, $\vec{u}=\vec{u}(\vec{x},t)$ is the velocity vector, $p=p(\vec{x},t)$ the fluid pressure, $\nu$ the kinematic viscosity (constant), $T_0$ a reference value for the temperature $T=T(\vec{x},t)$, $\vec{g}=(0,-g)$ the gravity acceleration pointing in the negative $y$ direction and $\alpha$ the thermal expansion coefficient (the Boussinesq approximation~\cite{chandrasekhar-hydrodynamic-2013,shyy-computational-2007} is assumed for the buoyancy forces). Regarding the equation for the thermal dynamics~\cref{eq:temperature-equation}, $\kappa$ is the thermal diffusivity, $L_f$ the latent heat of fusion and $C_p$ the specific heat of the substance. The field $\phi=\phi(\vec{x},t)$ is the \textit{liquid fraction}, defined as the relative volume occupied by the liquid phase in a certain position. $\phi$ depends on the local enthalpy $h$ which, in turn, depends on the local temperature field $T$~\cite{souayfane-melting-2018}. Before the phase transition process, no liquid is present in the computational domain and so $\phi=0$; when the phase transition is complete, only the liquid phase is present, with  $\phi=1$~\cite{shyy-computational-2007}. The relation between the liquid fraction $\phi$ and the other quantities is set by a smoothed step function~\cite{shyy-computational-2007}:
\begin{equation}
    \label{eq:phi-of-h}
    \phi(\vec{x},t)=
    \begin{cases}
        0                     & h<h_s           \\
        \frac{h(\vec{x},t)-h_s}{h_l-h_s} & h_s\le h\le h_l \\
        1                     & h>h_l
    \end{cases}
\end{equation}
where the local enthalpy $h (\vec{x},t)$ is written as a suitable linear combination of temperature and liquid fraction~\footnote{Notice that when discretizing equations, the liquid fraction in the rhs. of ~\cref{eq:explicit-enthalpy} is taken as the liquid fraction at the previous time step~\cite{proia-melting-2024}}
\begin{equation}\label{eq:explicit-enthalpy}
h(\vec{x},t)=C_pT(\vec{x},t)+L_f\phi(\vec{x},t).
\end{equation}
In the equations above, $h_s=C_pT_C$ is the solid phase enthalpy and $h_l=h_s+L_f$ the liquid phase enthalpy. The momentum dynamics and the thermal dynamics are two-way coupled via the buoyancy force in~\cref{eq:ns-momentum} and the advection velocity in~\cref{eq:temperature-equation}. Additionally, the bulk equations must be coupled with proper boundary conditions to describe the cell (walls and fin). From the fluid dynamics point of view, the walls and the fin are all solid boundaries, while from the thermal point of view we have the left wall and the fin as sources at constant temperature $T_H$ and the top, bottom and right walls are insulating:
\begin{equation}
    \label{eq:boundary-conditions-continuum}
    \begin{cases}
        \left.\vec{u}\right|_\text{wall,~fin}=0\\
        \left.T\right|_\text{left,~fin}=T_H=\text{const.}\\
        \left.\partial_{\vec{n}}T\right|_\text{top,~bottom,~right}=0~.
    \end{cases}
\end{equation}\\
The equations set~\cref{eq:ns-mass}-\cref{eq:temperature-equation} with the boundary conditions~\cref{eq:boundary-conditions-continuum} is solved with a lattice Boltzmann scheme~\cite{guo-coupled-2002,huber-lattice-2008}. Other schemes able to solve the fluid-thermal equations are known from the literature~\cite{wen-Development-2021,wang-modified-2016}, but we opted for the LBM scheme that has been extensively validated in our earlier study~\cite{proia-melting-2024}, where we also discuss the advantages of such approach. The interested reader is referred to this publication (and references therein) for details on the numerical methodology.\\
To study the dynamics of melting, it is useful to introduce the non-dimensional  Rayleigh Number $\ra$:
\begin{equation}\label{eq:rayleigh-number}
\ra=\frac{g\alpha L^3}{\nu\kappa}(T_H-T_C) \, ,
\end{equation}
representing the ratio between buoyancy forces and diffusion/dissipation and the Stefan Number $\ste$:
\begin{equation}\label{eq:stefan-number}
\ste=\frac{C_p}{L_f}(T_H-T_C) \, ,
\end{equation}
representing the ratio between sensible and latent heat. We systematically varied the values of $\ra$ and $\ste$ in order to assess their effect on the PCM dynamics, while the Prandtl Number $\pr=\nu/\kappa$ is kept fixed in all simulations.\\
The insertion of the fin effectively modifies the heat source surface and the volume of the substance with respect to the finless cell, so to correctly monitor the actual speed-up it is useful to define an observable corresponding to the space integral of the liquid fraction normalized to the PCM volume $V_{\text{PCM}}=L^2-S$
\begin{equation}
    \label{eq:phi-star}
    \phist(t)=\frac{\sum_{\vec{x}}\phi(\vec{x},t)}{V_{\text{PCM}}}\le1
\end{equation}
By definition, $\phist(0)=0$ and when the PCM has completely melted $\phist=1$. From this point on, we shall refer to $\phist$ as ``liquid fraction'' for the sake of brevity.
\section{Results and discussion}\label{sec:results}
We perform simulations that vary both the fin parameters $\hat l,\hat h$ and the non-dimensional numbers $\ste,\ra$. The ranges are as follows:
$\hat l\in\left\{0.1,0.2,0.4,0.5\right\}$,
$\hat h\in\left\{0.1,0.2,0.5,0.8,0.9\right\}$,
$\ste\in\left\{0.1,1,10\right\}$,
$\ra\in\left\{10^5,10^6,10^7\right\}$ for a total of $180$ finned simulations plus $9$ finless simulations. The other parameters are fixed at $\pr=1,\,L=250,\,T_H=1,\,T_C=0,\,T_0=0.5$. All the simulations were performed until the melting time $t_m$, i.e. the time when the melting process is complete.\\
In~\cref{fig:time-evo} we report a comparison of the time evolution of the temperature field $T(\vec{x},t)$ for two systems: a finless one (upper plots) and a finned one (lower plots) with the same non-dimensional numbers $\ra=10^7$ and $\ste=10$. Time grows from left to right. It is evident how the presence of the fin translates in a faster melting speed. As also stated later in this article, this is due to the development of convective plumes between the fin and the upper wall (visible in the lower plots) that are the major vectors for heat transfer, allowing heat to reach the whole domain faster with the development of a flow current spanning the whole system (third figure in the lower plots). The systems that develop the biggest plumes are also the fastest.\\
We can further investigate how the position and length of the fin influence the melting time: in~\cref{fig:fin-pos-comparison} we plot snapshots of the temperature field for finned systems with different  realizations of $\hat{l}$ and $\hat{h}$ for $t=\num{1.3E4},\,\ra=10^6,\,\ste=1$; the corresponding liquid fraction $\phi^*$ is also indicated. We can see that the dynamics triggered by the longer, lower-positioned fins is substantially different from the others, with the presence of convective plumes that provide a speed-up in the melting rate. On the contrary, very short or very high fin have a very similar dynamics, that closely resemble the dynamics of the finless cell (see~\cref{fig:time-evo}). In~\cref{fig:phistar-h} we provide a more detailed analysis by plotting the data for $\phist$ in the early stages of melting at time $t=\num{1.3E4}$, changing $\ra$ between~\cref{subfig:comparison-a} and~\cref{subfig:comparison-b}, while $\ste$ increases from left to right. Notice that the left plot of~\cref{subfig:comparison-b}, being the one with highest $\ra$ (lower $\kappa$) and lowest $\ste$ (higher $L_f$), is the one closest to an actual PCM system.  Again, it can be seen that the insertion of the fin leads to a general improvement of the melting rate and it is evident how -- for fixed $\ra$ and $\ste$ -- the general trend is to have a larger liquid fraction for lower and longer fins. From this figure we can also deduce that, at fixed $\ste$, higher $\ra$ means better performance. On the other hand, at least in these early stages of the dynamics, we can see that $\ste$ seems to have more influence on the speed-up provided by the fin compared to $\ra$, since lower Stefan means slower melting by itself. Despite referring to a single time in the early stages of the dynamics, we observe in the left plots in~\cref{fig:fin-pos-comparison} a non monotonic behavior of $\phist$ as a function of $\hat{h}$: this fact will be further discussed lated in the context of the global melting time.\\
Both from the snapshots reported in~\cref{fig:time-evo} and~\cref{fig:fin-pos-comparison} as well as from the liquid fraction $\phi^*$ at a given time $t$ analyzed in~\cref{fig:phistar-h} we can get a clear indication that the presence of the fin influences in a non-trivial way the melting dynamics with a net impact on the values of the non-dimensional numbers $\ste$ and $\ra$; however, we hasten to remark that the difference in total melting time is quite large between simulations, hence it is difficult to find a selected time $t$ which is fully representative of the trend of the melting for all the different systems, without misrepresenting systems that appear to be slower in earlier times but experience speed-ups at later stages. For these reasons, we conducted a systematic analysis on the total melting time $t_m(\hat{l},\hat{h},\ste,\ra)$ normalized to the total melting time of the corresponding finless cell $t_m(0,0,\ste,\ra)$
\begin{equation}
    \label{eq:norm-melting-time}
    \hat{t}_m(\hat{l},\hat{h},\ste,\ra)=\frac{t_m(\hat{l},\hat{h},\ste,\ra)}{t_m(0,0,\ste,\ra)}\,.
\end{equation}
The lowest values of $t_m$ indicate the optimal combination of parameters, providing the best boost for melting. In~\cref{fig:metling-time-comparison} we systematically analyze $\hat{t}_m$ at changing $\hat{h}$ and $\hat{l}$ for different realizations of $\ste$ and $\ra$. Here we can see that, for all values of $\ra$ and $\ste$, the lowest values of $\hat{t}_m$ (green, close to $0.5$) are generally obtained for the longest and lowest fin. This result well agrees with previous results in the literature~\cite{wu-phase-2021}, although the analysis presented in~\cite{wu-phase-2021} focuses on the coupling of melting and solidification and not on the influence of $\ra$ and $\ste$ which are kept fixed at $\ra\approx\num{4.65e5},\,\ste\approx\num{3.07e-1}$ (these values are computed based on the lauric acid used in~\cite{wu-phase-2021}). The authors in~\cite{gao-lattice-2017}, instead, report the fin's vertical position to have no influence on the melting; however, the study in~\cite{gao-lattice-2017} also features a porous medium that, while improving conduction, hinders the formation of big convective rolls~\cite{wang-critical-2022,feng-porescale-2015}. A similar conclusion on the role of the fin's vertical position can be found in~\cite{jourabian-simulation-2012}, where the authors simulate a finned system with no porous medium, investigating the effects of position and length of the fin, finding only the latter to be relevant towards melting speed; however, the authors chose to measure the melting speed in term of the average melting front (i.e. the boundary between solid and liquid substance) as a function of $t$; this measure is actually offset by the presence of the fin. For this reason, in the present work, we decided to compute the melting time as the time when the whole cell has molten, resulting in $\phist=1$. To better appreciate the modulation in $\hat{t}_m$ induced by a change in $\ra$ and $\ste$, in~\cref{fig:metling-time-comparison} we also draw contour lines for some selected values of $\hat{t}_m$ in the range $0.8-0.97$. From these contour lines we can better appreciate the impact of a change in $\ra$ and $\ste$: while the variation of the second one does not seem to have a strong effect on the system, varying the first greatly influences the shape of the contours, promoting an enlargement of the region where $\hat{t}_m\le0.8$. High $\ra$, indeed, promote convection in the system, thus favoring buoyancy forces rather than diffusion/dissipation. On the other hand, $\ste$ simply regulates how fast the substance conducts heat with respect to how much heat it absorbs during the transition, and since convection is the main mechanism of heat transfer at long time scales~\cite{feng-porescale-2015}, conduction does not have a significant effect on the long term dynamics other than a little increase in melting time for lower Stefan due to larger latent heat (see the middle plots in~\cref{fig:metling-time-comparison}). Notice that the drop in values of $\phist(t)$ between high and low values of $\ste$ observed in~\cref{fig:phistar-h} is probably due to the fact that in the early stages and for lower $\ste$ there is not sufficient molten substance for convection to exert its effects and grant a significant speed-up. This is also shown by the fact that for higher $\ste$ some system can reach $\phist\approx1$ even in these early stages (see the right plot of~\cref{subfig:comparison-a}).\\
For more quantitative results on the melting time, we can refer to~\cref{fig:t0-an}: here, we plot $\hat{t}_m(\hat{h})$ for different values of $\ra,~\ste$; different $\hat{l}$ are reported with different line colors and symbols. We also indicate the global~\footnote{i.e. computed over all the simulations, not only those with the same non-dimensional numbers.} maximum and minimum for $\hat{t}_m$ with a shaded area. From these figures we can conclude that even with the single fin, the speed-up in melting can be as high as $60\%$ and that there is some non-trivial behavior, in that a local minimum at changing $\hat{h}$ appears for the largest $\ste$ and $\ra$.
To delve deeper, we investigate in~\cref{fig:vorticity-comparison} the velocity magnitude $|\vec{u}(\vec{x},t)|$ in the early stages of the melting process ($t=2\cdot10^3$) for three different fin heights corresponding to $\hat{h}=0.1,~0.2,~0.5$ at fixed $\hat{l}=0.5$, $\ra=10^7$, $\ste=10$. We also report the vorticity field $w_z=\partial_x u_y-\partial_y u_x$ in a region around the fin along with a vector plot of $\vec{u}(\vec{x},t)$. We also analyze in~\cref{fig:vorticity-plot} the vorticity field in the late stage of the melting process ($t=1.4 \cdot 10^4$). In~\cref{fig:phi-evo}, instead, we report the time evolution of $\phist(t)$. Taken all together, these figures help in elucidating the mechanism at the core of the non-monotonic behavior of $\hat{t}_m$ as a function of $\hat{h}$ observed in~\cref{fig:t0-an}. In the early stages of melting (see~\cref{fig:vorticity-comparison}), we observe the formation of small rolls on the top of the fin in all cases. However, by increasing the fin height, a convective motion is promoted in the sub-cell between the fin and the lower wall; such a convective motion is suppressed for the smallest $\hat{h}$, as seen from the weaker vorticity field. As a result, in the early stages of the melting process the largest heights are faster (see~\cref{fig:phi-evo}). This property, however, is not preserved at all times; indeed, in the late stages of melting, an increase of $\hat{h}$ comes with a reduction of the sub-cell height between the fin and the upper wall, thus reducing the intensity and the spatial extension of the convective motion in such a region and hindering the formation of a macroscopical flow current spanning the whole domain (see~\cref{fig:vorticity-plot}).
As a result, the higher fin slows down the melting process, reducing the heat transfer compared to  the cases with smaller $\hat{h}$. Summarizing, based on the observations in the late stages of the dynamics (see~\cref{fig:vorticity-plot}), one would be tempted to associate the fastest dynamics (smallest $\hat{t}_m$) to the smallest $\hat{h}$; however, for small $\hat{h}$, the dynamics is further influenced by the convection under the fin; its absence for very small heights results in a slower melting process. Thus, we find that an optimal height \emph{exists}, at which the overall melting process is faster, implying a minimum in $\hat{t}_m$.\\

\section{Conclusions}\label{sec:conclusions}

Given the importance of Phase Change Materials (PCMs) for the energy transition and the various interesting aspects of their heat transfer optimization~\cite{togun-critical-2024,wang-critical-2022,proia-melting-2024}, in this study we performed numerical simulations to assess the influence of a fin insertion in a PCM cell, by systematically investigating the impact of the fin geometry (given by its normalized height $\hat{h}$ and length $\hat{l}$) on the melting time normalized to the finless configuration, $\hat{t}_m$. Importantly, we conducted the study by varying both the Rayleigh number $\ra$ - which compares convective to dissipative/diffusive effects - and the Stefan number $\ste$ - relating conductive to latent heat. Our results confirm that the insertion of a fin generally shortens PCM melting time. Such an improvement, however, depends on $\ste$ and $\ra$, with $\ra$ having a stronger effect on the total melting time. In particular, for the largest $\ra$ analyzed, we found a non-monotonic behavior of $\hat{t}_m$ as a function of $\hat{h}$, for fixed $\ste$ and $\hat{l}$. Such non-monotonic behavior can be rooted in the presence of an optimal height that favors the convective circulations below the fin in the early stages of the melting dynamics, still promoting large scale convective structures above the fin in the late stages. This evidence points to interesting topological implications, which can be the object of future systematic analyses~\cite{guo-mechanisms-2025}. As a further, future development, since the present study was conducted in a 2D cell, varying the length and height of single fin, the extension to 3D, as well as the evaluation of system performance in the presence of multiple fins with different orientations would provide interesting insights for the PCM-based TES field~\cite{sharifi-enhancement-2011}.
\begin{minipage}{\linewidth}
 \vspace*{5mm}
\subsection*{Acknowledgments}
The authors wish to acknowledge the support of the National Center
for HPC, Big Data and Quantum Computing, Project CN\_00000013
- CUP E83C22003230001, Mission 4 Component 2 Investment 1.4,
funded by the European Union - NextGenerationEU.\\
G.F. and P.P. wish to acknowledge the support of Project PRIN 2022F422R2 - CUP E53D23003210006, financed by the European Union – Next Generation EU and the support of the Office of Naval Research, United States with Drs. J. Dibelka and E. McCarthy as program managers (Grant N62909-24-1-2072).\\
G.F. also wishes to acknowledge the support of Project PRIN PNRR P202298P25 - CUP E53D23016990001, financed by the European Union – Next Generation EU, and Project ECS 0000024 Rome Technopole, - CUP B83C22002820006, NRP Mission 4 Component 2 Investment 1.5,  Funded by the European Union - NextGenerationEU.
\end{minipage}

\begin{minipage}{\linewidth}
\begin{figure}[H]
    \centering
    \includegraphics[width=1\linewidth]{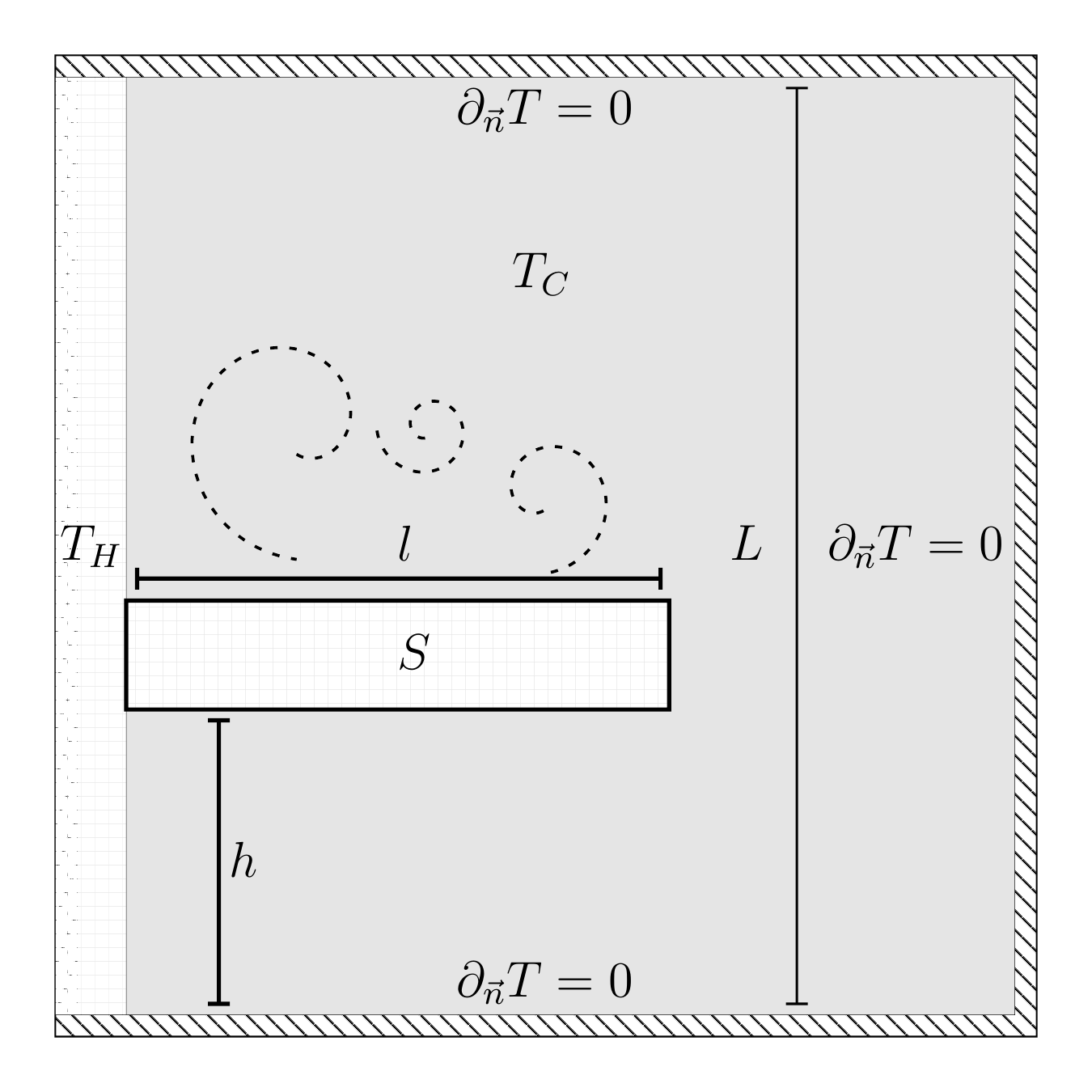}
    \caption{Sketch of PCM cell geometry. The enclosure consists of a square cell with a side length of $L$. On the left side there is a source at fixed temperature $T_H$, from which a fin of fixed surface area $S$ and variable length $l$ protrudes at height $h$. The fin is an heat source at temperature $T_H$. Its height and length are normalized to the side length as $\hat{l}=l/L$ and $\hat{h}=h/L$. The three other sides are insulating wall for which $\partial_{\bar{n}}T=0$, with $\bar{n}$ being the normal to the wall. The enclosure is filled with the solid substance at the melting temperature $T_C$. \label{fig:cell-geometry}}
\end{figure}
\end{minipage}

\clearpage
{
\begin{figure*}
    \centering
    \includegraphics[width=\linewidth]{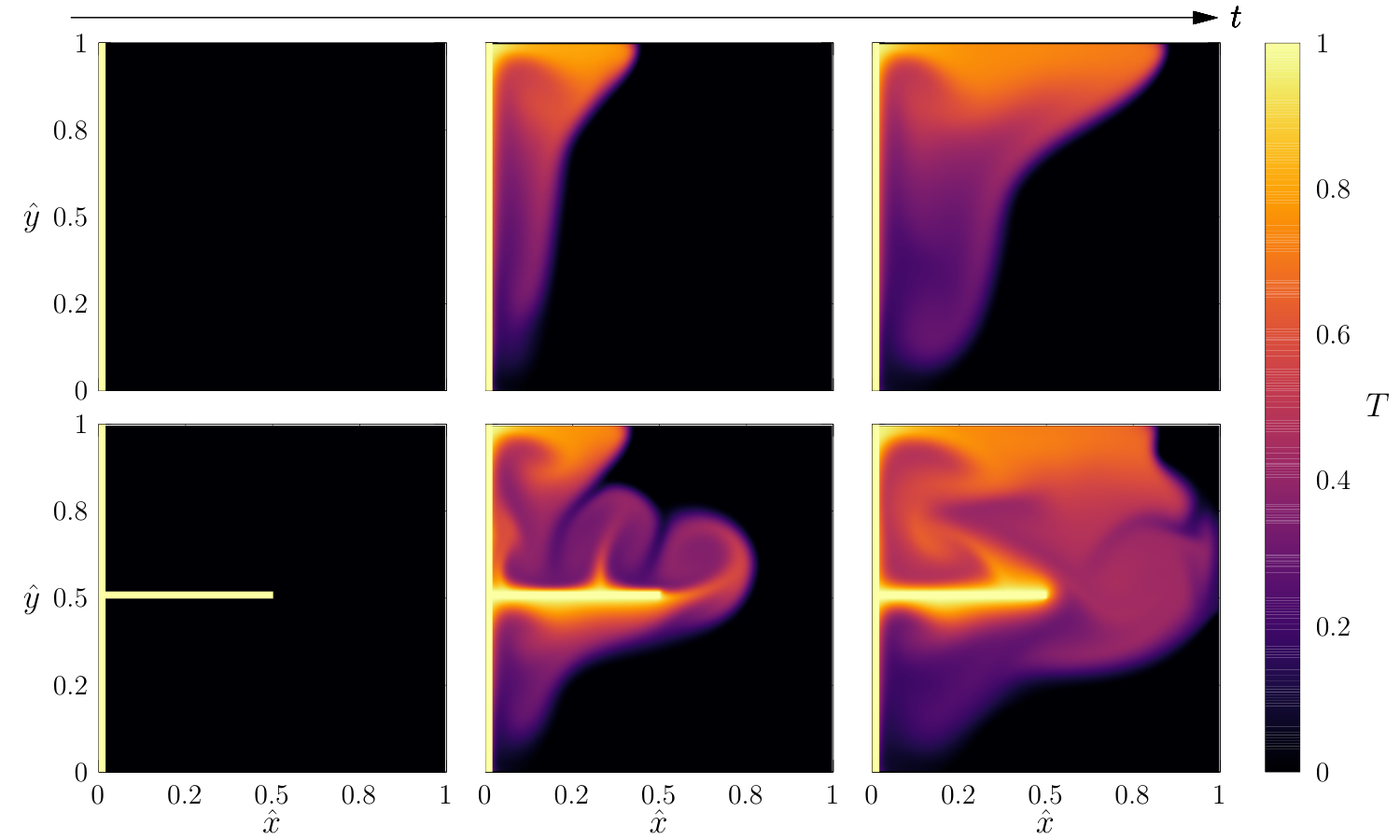}
    \caption{Comparison of the time evolution of the temperature field $T(\vec{x},t)$ for non-finned (upper plots) and finned (lower plots) cells with $\operatorname{Ra}=10^7,~\operatorname{St}=10$. Time increases from left to right. The space coordinates are normalized to the size of the cell $L$ as $\hat{x}=x/L$ and $\hat{y}=y/L$.\label{fig:time-evo}}
\end{figure*}
\begin{figure*}
    \centering
    \includegraphics[width=\linewidth]{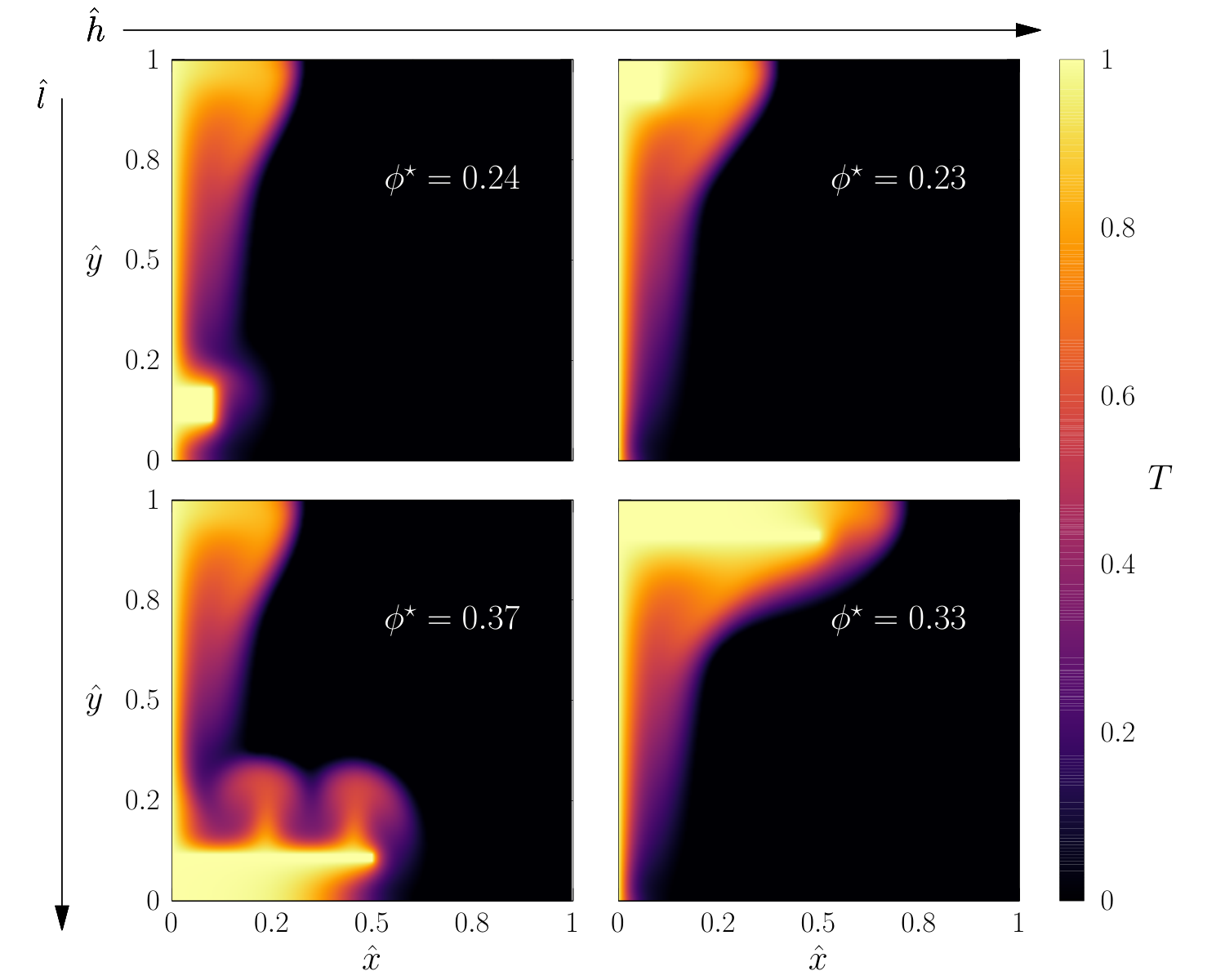}
    \caption{Comparison of the temperature field $T(\vec{x},t)$ and liquid fraction $\phist(t)$~(see~\cref{eq:phi-star}) for $\operatorname{Ra}=10^6,~\operatorname{St}=1$ for various fin height/length configurations at time $t=\num{1.3E4}$. $\hat{h}$ increases from left to right, while $\hat{l}$ increases from top to bottom.\label{fig:fin-pos-comparison}}
\end{figure*}
\begin{figure*}
    \centering
    \begin{subfigure}{\linewidth}
    \caption{\large$\ra=10^5$}
    \label{subfig:comparison-a}
    \includegraphics[width=\linewidth]{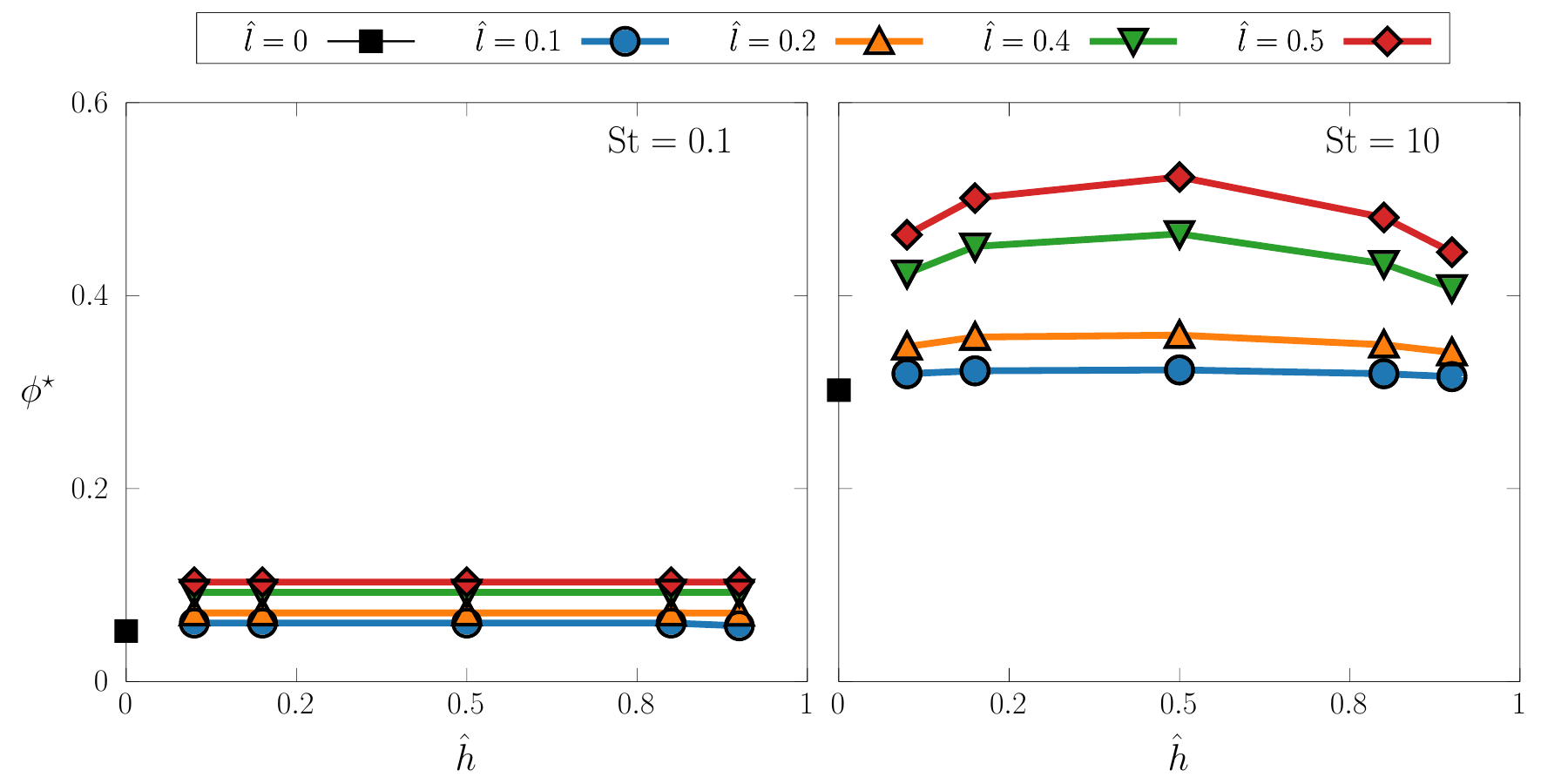}
    \end{subfigure}
    \begin{subfigure}{\linewidth}
    \caption{\large$\ra=10^7$}
    \label{subfig:comparison-b}
    \includegraphics[width=\linewidth]{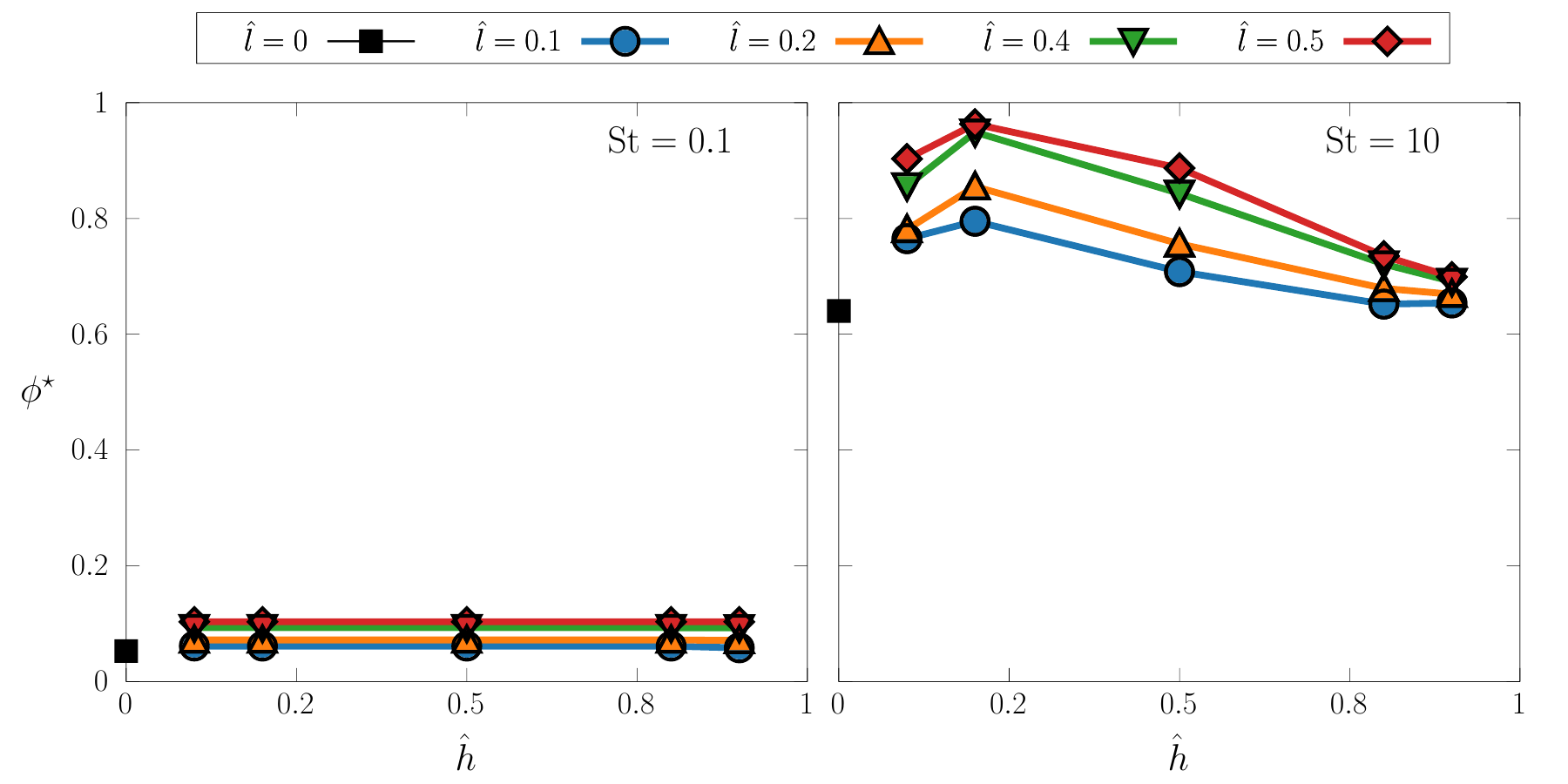}
    \end{subfigure}
    \caption{We report the liquid fraction $\phist$~(see~\cref{eq:phi-star}) as a function of $\hat{h}$ at time $t=\num{1.3E4}$, for different values of $\hat{l},~\ra,~\ste$. $\ra$ changes between sub-figures in panels (a) and (b), while $\ste$ increases from left to right. The different lengths are shown with different symbols and colors; the black square represents the value for the case without the fin.\label{fig:phistar-h}}
\end{figure*}
\begin{figure*}
    \centering
    \includegraphics[width=\linewidth]{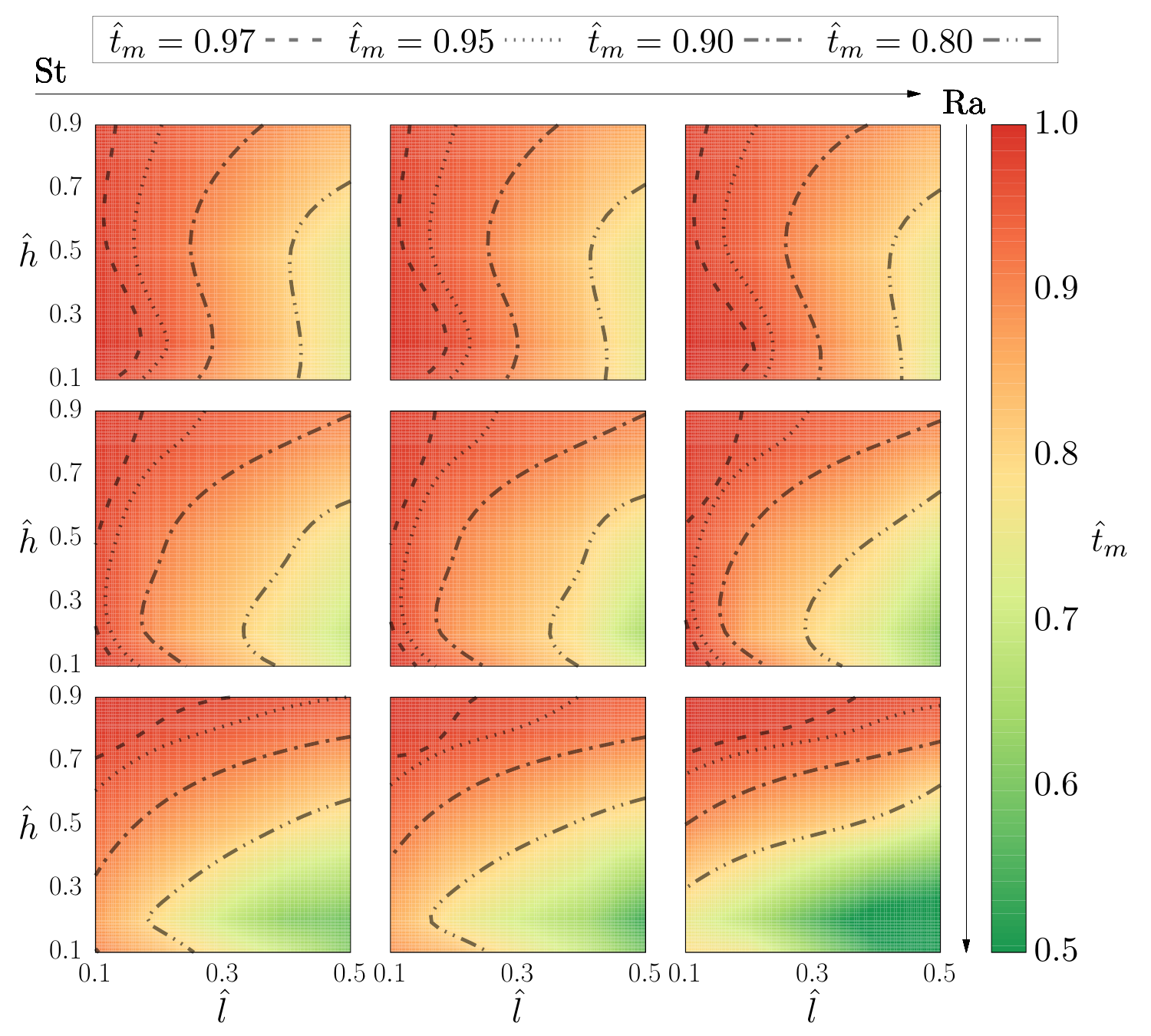}
    \caption{Plot of the normalized melting time $\hat{t}_m$ (see~\cref{eq:norm-melting-time}) at changing $\hat{l},~\hat{h},~\ra~,\ste$.  $\ste$ increases from left to right, while $\ra$ increases from top to bottom. We also show contour lines for some selected values of $\hat{t}_m$ in the range $0.8-0.97$.}
    \label{fig:metling-time-comparison}
\end{figure*}
\begin{figure*}
    \centering
    \includegraphics[width=\linewidth]{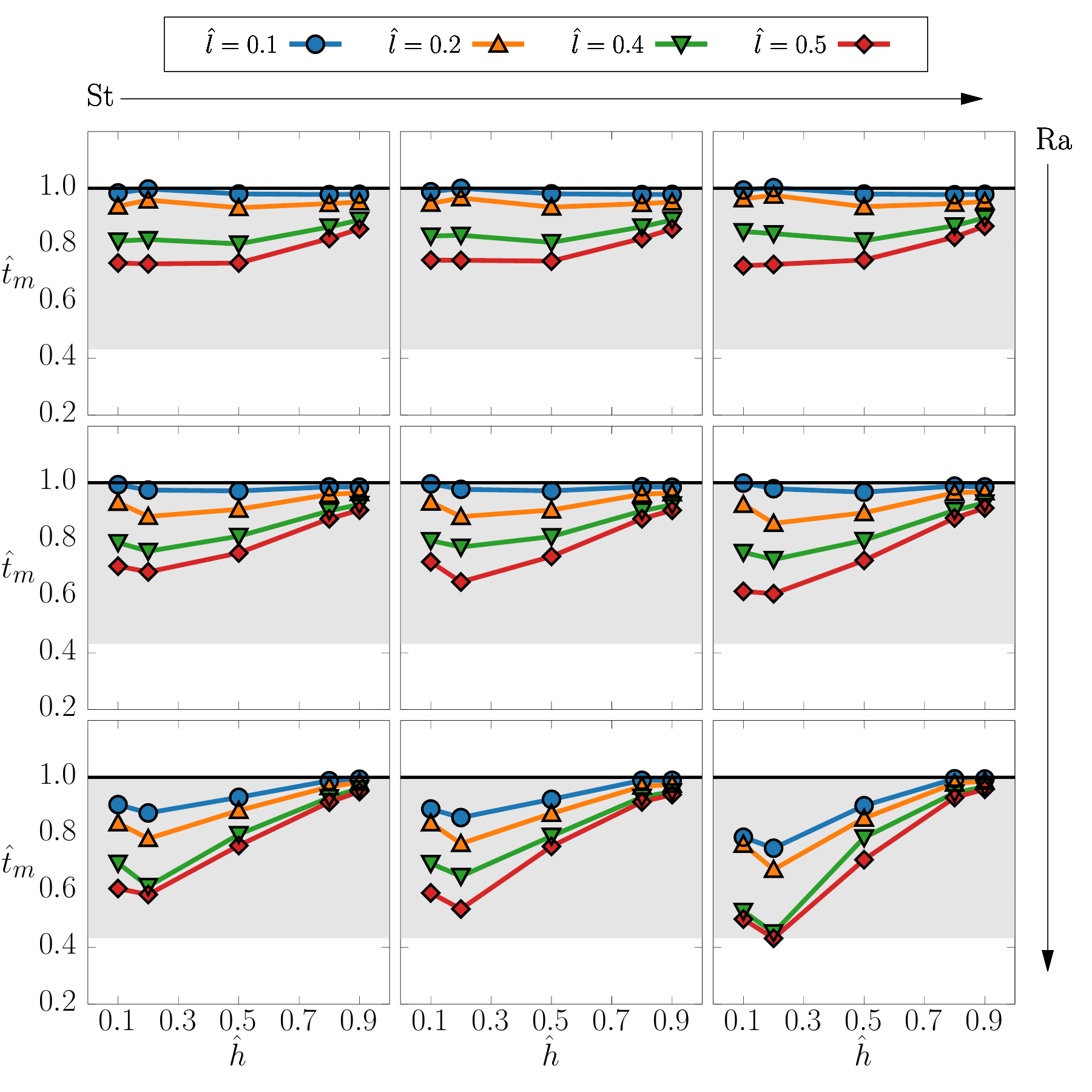}
    \caption{Plot of the normalized melting time $\hat{t}_m$ (see~\cref{eq:norm-melting-time}) as a function of $\hat{h}$, for different values of $\hat{l}$ and different realizations of $\ra$ and $\ste$. $\ste$ increases from left to right, while $\ra$ increases from top to bottom. The different values of $\hat{l}$ are shown with different symbols and colors; the black line at $\hat{t}_m=1$ corresponds to the finless case. The shaded area is drawn in correspondence of the global maximum and minimum for all $\hat{t}_m$.\\ \label{fig:t0-an}}
\end{figure*}

\begin{figure*}
    \centering
    \includegraphics[width=\linewidth]{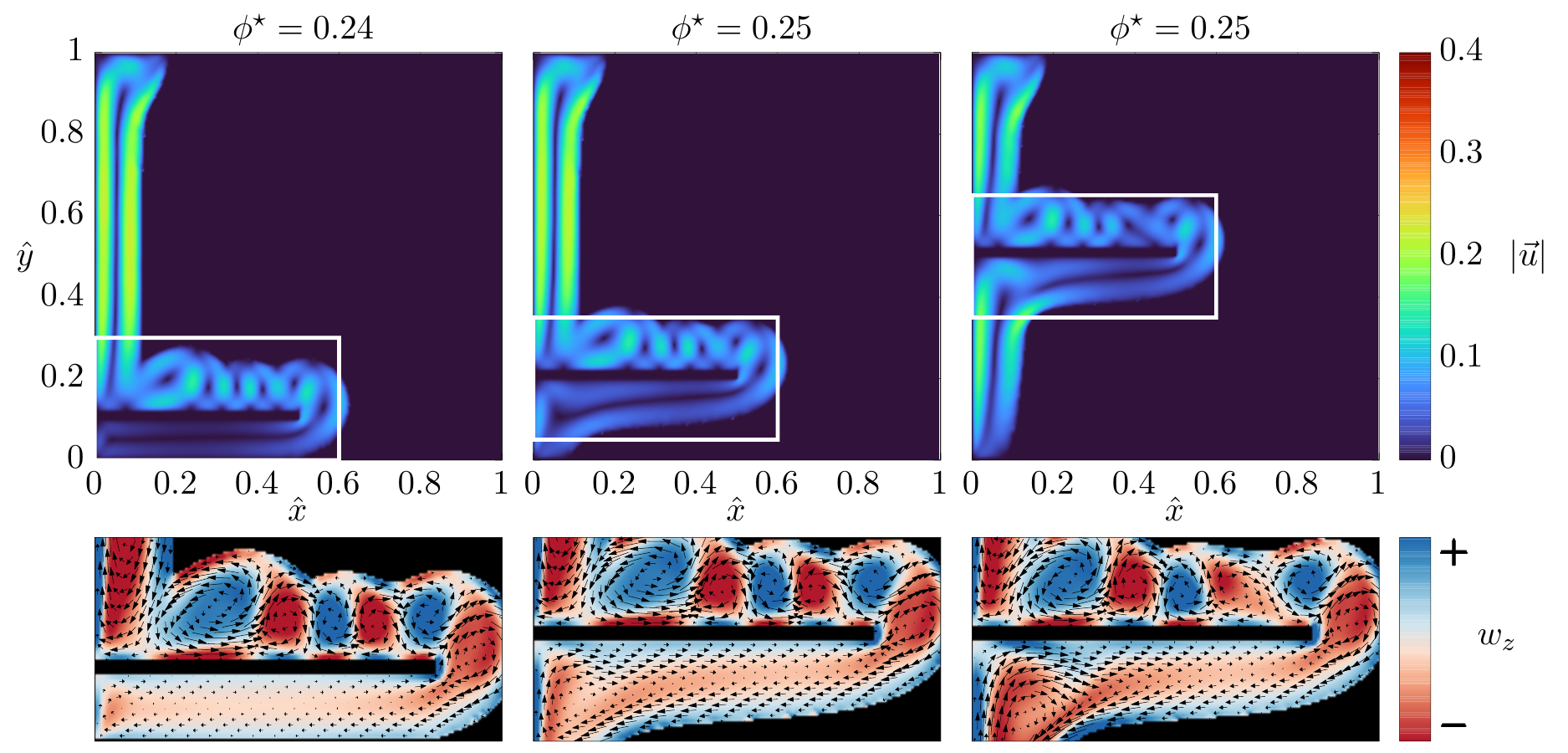}
    \caption{Comparison of the velocity magnitude $\left| \vec{u}(\vec{x},t) \right|$ at time $t=2\cdot10^3$ for three different fin heights corresponding to $\hat{h}=0.1,~0.2,~0.5$ at fixed $\hat{l}=0.5$, $\ra=10^7$, $\ste=10$. The corresponding liquid fraction $\phist$ is indicated. In the lower part of the figure a snapshot of the vorticity $w_z(\vec{x},t)$ in proximity of the fin is provided along with a vector plot of $\vec{u}(\vec{x},t)$.\label{fig:vorticity-comparison}}
\end{figure*}

\begin{figure*}
    \centering
    \includegraphics[width=\linewidth]{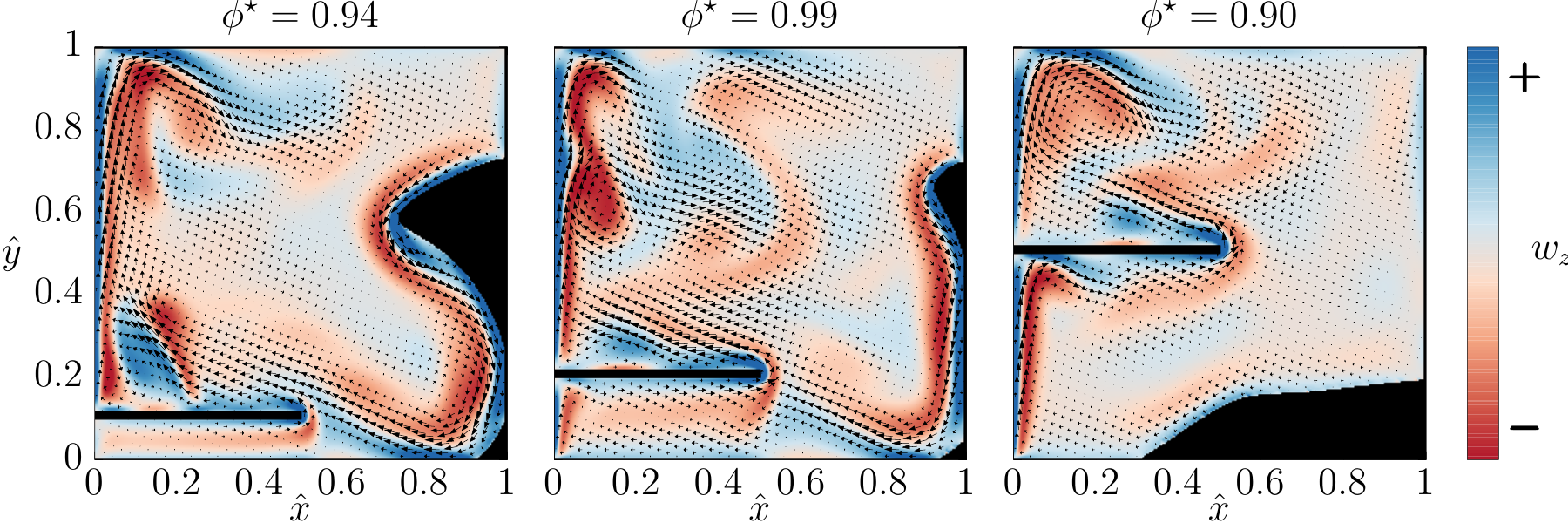}
    \caption{Snapshot of the vorticity $w_z(\vec{x},t)$ along with a vector plot of $\vec{u}(\vec{x},t)$ at $t=1.4\cdot10^4$ for three different fin heights corresponding to $\hat{h}=0.1,~0.2,~0.5$ at fixed $\hat{l}=0.5$, $\ra=10^7$, $\ste=10$. The corresponding liquid fraction $\phist$ is indicated.\label{fig:vorticity-plot}}
\end{figure*}

\begin{figure*}
    \centering
    \includegraphics[width=\linewidth]{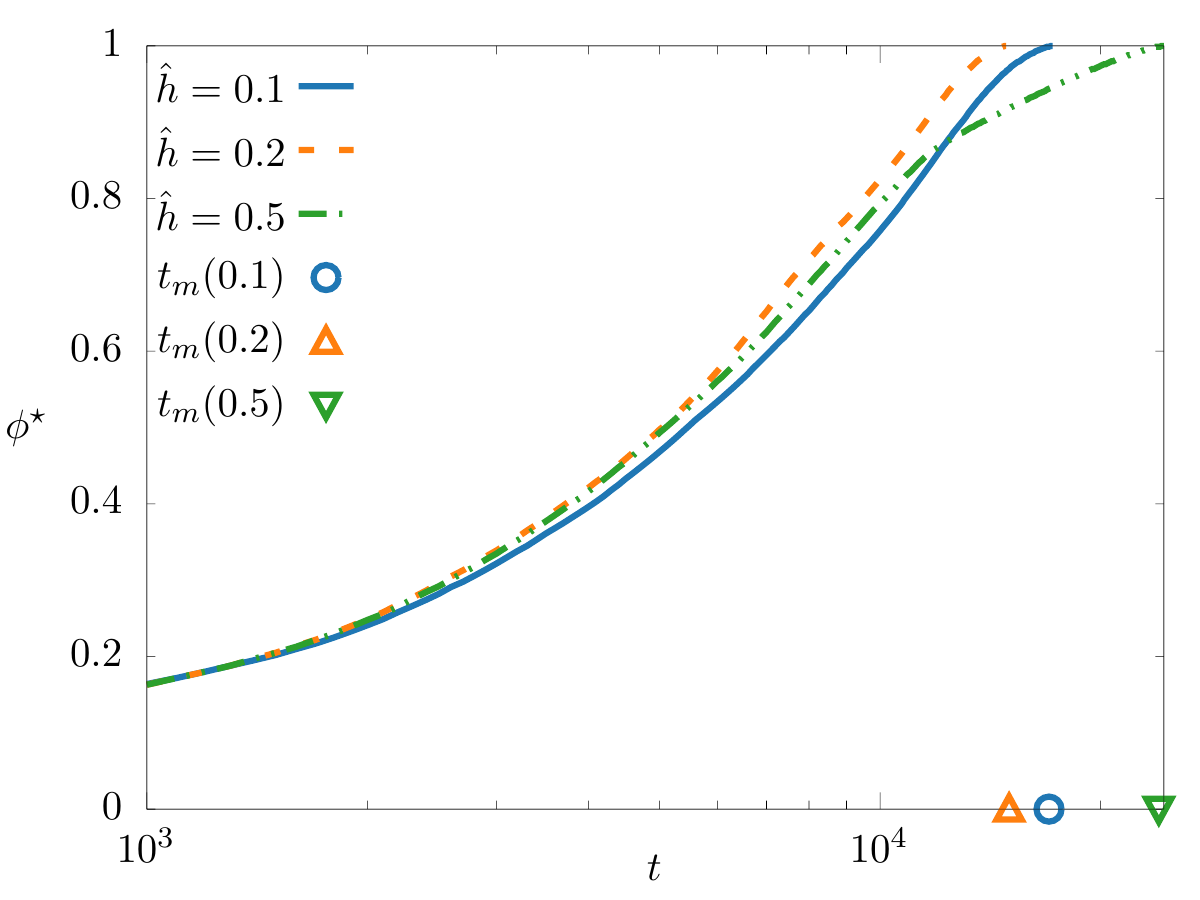}
    \caption{We report the liquid fraction $\phist$~(see~\cref{eq:phi-star}) for the three cases examined in \cref{fig:vorticity-comparison}.  Points on the $x$ axis mark the three different melting times $t_m$.\label{fig:phi-evo}}
\end{figure*}

\FloatBarrier
\bibliographystyle{elsarticle-num}
\bibliography{PCM.bib,LatticeBoltzmann.bib,FluidDynamics.bib}
}
\end{document}